\documentstyle[prl,aps,multicol,amssymb,epsf]{revtex} 

\tighten

\begin{document}
\draft
\title{Two-scale localization in disordered wires in a magnetic field.}
\author{A.~V.~Kolesnikov$^{1}$and K.~B.~Efetov$^{1,2}$}
\address{$^{1}$Fakult\"{a}t f\"{u}r Physik und Astronomie, Ruhr-Universit\"{a}t\\
Bochum, Universit\"{a}tsstr. 150, Bochum, Germany\\
$^{2}$L.D. Landau Institute for Theoretical Physics, Moscow, Russia}
\date{\today{}}
\maketitle

\begin{abstract}
Calculating the density-density correlation function for disordered wires,
we study localization properties of wave functions in a magnetic field. The
supersymmetry technique combined with the transfer matrix method is used. It
is demonstrated that at arbitrarily weak magnetic field the far tail of the
wave functions decays with the length $L_{{\rm cu}}=2L_{{\rm co}}$, where $%
L_{{\rm co}}$ and $L_{{\rm cu}}$ are the localization lengths in the absence
of a magnetic field and in a strong magnetic field, respectively. At shorter
distances, the decay of the wave functions is characterized by the length $%
L_{{\rm co}}$. Increasing the magnetic field broadens the region of the
decay with the length $L_{{\rm cu}}$, leading finally to the decay with $L_{%
{\rm cu}}$ at all distances. In other words, the crossover between the
orthogonal and unitary ensembles in disordered wires is characterized by two
localization lengths. This peculiar behavior must result in two different
temperature regimes in the hopping conductivity with the boundary between
them depending on the magnetic field.
\end{abstract}


\begin{multicols}{2} 

Disordered systems have been under intensive study for several decades but
only recently strong localization at moderate disorder in 1D wires has
become the subject of a systematic experimental investigation \cite{Exp}.
The authors of Ref.\cite{Exp} studied electron transport in
submicrometer-wide wires fabricated from Si $\delta $-doped GaAs. A large
number of the wires were connected in parallel and the conductivity of the
entire system was measured as a function of temperature $T$ and magnetic
field $H$ applied perpendicular to the wires. The activation temperature
dependence observed allowed to demonstrate the exponential localization and
extract the dependence of the localization length $L_{c}$ on the magnetic
field. It was found that in a strong magnetic field the localization length
is twice as large as in the absence of the magnetic field.

Localization due to quantum interference is a fundamental property of
disordered one-dimensional (1D) systems. Mott and Twose \cite{MT} predicted
that at arbitrary weak disorder all states of a disordered chain had to
become localized, which has been proven later \cite{Ber}. Thouless \cite{T1}
realized that all wave functions had to be localized also in thick wires.
Such wires are more interesting because electrons can diffuse at distances
exceeding the mean free path $l$ but get localized on the localization
length $L_{{\rm c}}=l\left( p_{0}^{2}S\right) $, where $p_{0}$ is the
Fermi-momentum and $S$ is the cross-section of the wire. Explicit
calculations confirmed this picture for thick wires \cite{EL,efetov} as well
as for systems of coupled chains \cite{D}. In contrast to the chains, the
electron motion in thick wires is sensitive to an external magnetic field. A
remarkable effect, namely, doubling of the localization length when applying
a strong magnetic field $H$ was predicted for such systems \cite{EL,D}. It
is this doubling that has been observed in Ref.\cite{Exp}. Different aspects
of the localization in wires as well as connections to random matrix
theories were discussed later\cite{kram,lerner,beenak,book,pich}.

Surprisingly, little attention has been paid to the crossover between the
limits of zero and strong magnetic fields $H$ (between the orthogonal and
unitary ensembles). Study of this crossover is important, however, because
the localization length can be measured experimentally. Except for an
interpolation formula for the localization length suggested in Ref. \cite{FR}
and a numerical study of Ref. \cite{pich}, practically no attempt has been
made to describe the crossover. Apparently, the absence of attention has
been due to a common belief that the crossover is simple and the only thing
that can happen is that the localization length 
changes smoothly between its value $L_{{\rm co}}$ at zero field and $L_{{\rm %
cu}}=2L_{{\rm co}}$ at a strong magnetic field. Such a scenario was used,
for example, by the authors of Ref. \cite{Exp} in their attempt to fit the
data on the interpolation curve of Ref. \cite{FR}, suggested basically to
describe the smooth change of the localization length.

In this Letter we present results of an analytical study of a correlation
function describing the decay of wave functions in disordered quantum wires.
In contrast to the one-scale picture, the behavior of the wave functions at
finite magnetic fields turns out to be more complicated. The most striking
picture of this decay is that, even at very weak magnetic fields, the far
tail of the wave functions falls off with the length $L_{{\rm cu}}$, whereas
at smaller distances the decay is governed by the length $L_{{\rm co}}$. The
larger length $L_{{\rm cu}}$ does not depend on the strength of the magnetic
field. Increasing the magnetic field results in a broadening of the
asymptotic region until the entire wave function starts decaying with the
length $L_{{\rm cu}}$.

The behavior of localized wave functions can be well described by the
correlation function $p_{\infty }\left( r\right) $%
\begin{equation}
p_{\infty }\left( r\right) =\left\langle \sum_{\alpha }\left| \psi _{\alpha
}\left( 0\right) \right| ^{2}\left| \psi _{\alpha }\left( r\right) \right|
^{2}\delta \left( \varepsilon -\varepsilon _{\alpha }\right) \right\rangle
\;,  \label{0}
\end{equation}
where $\psi _{\alpha }\left( r\right) $ and $\varepsilon _{\alpha }$ are the
eigenfunction and eigenenergy of a state $\alpha $, $r>0$ is the coordinate
along the wire,\ and the angle brackets stay for averaging over disorder.

The function $p_{\infty }\left( r\right) $ is important not only for a
theoretical description of localized wave functions but it also determines
directly the hopping conductivity at low temperature. According to results
obtained quite long ago \cite{kurk,larkin}, the hopping conductivity of
one-dimensional chains and wires is described by a simple formula 
\begin{equation}
\sigma =\sigma _{0}\exp \left( -T_{0}/T\right) \text{, \ \ \ }T_{0}\sim
\left( \nu _{1}L_{{\rm c}}\right) ^{-1}\text{, \ \ \ }\nu _{1}=\nu S\; ,
\label{01}
\end{equation}
where $\nu $ is the density of states, $\nu _{1}$ is the one-dimensional
density of states, $S$ is the cross-section of the wire and $L_{{\rm c}}$ is
a localization length that can be extracted from the function $p_{\infty
}\left( r\right) $, Eq.~(\ref{0}). The pre-factor $\sigma _{0}$ depends on
parameters of the model. Remarkably, even the Coulomb interaction does not
change the temperature dependence, Eq.~(\ref{01}), entering $\sigma _{0}$
only \cite{larkin}. Equation (\ref{01}) was used in Ref. \cite{Exp} for
extracting the localization length from the temperature dependence of the
conductivity.

The correlation function $p_{\infty }\left( r\right) $ can be found using
the supersymmetry technique \cite{efetov,book}. 
Let us first discuss the final result in a weak field which reads 
\begin{eqnarray}
p_{\infty }(r)\simeq\frac{1}{4\sqrt{\pi }L_{{\rm co}}}\left( \frac{\pi ^{2}}{%
8}\right) ^{2}\left( \frac{4L_{{\rm co}}}{r}\right) ^{3/2}  \nonumber \\
\times\left[ \exp \left( -\frac{r}{4L_{{\rm co}}}\right) +\alpha X^{2} \ln^2
X \exp \left( -\frac{r}{4L_{{\rm cu}}}\right) \right] \;  \label{02}
\end{eqnarray}
where $L_{{\rm co}}=\pi \nu _{1}D_{0}$ is the localization length for the
orthogonal ensemble, $L_{{\rm cu}}=2L_{{\rm co}}$ is the localization length
for the unitary ensemble, $D_{0}$ is the classical diffusion coefficient,
and $\alpha$ is a positive constant of order unity. The parameter $X=2\pi
\phi /\phi _{0}$ describes the crossover between the orthogonal and unitary
ensemble, $\phi _{0}=hc/e$ is the flux quantum, and $\phi =2HL_{{\rm co}%
}\left\langle y^{2}\right\rangle _{\sec }^{1/2}$ is a characteristic
magnetic flux through an area limited by the localization length. The
coordinate $y$ is perpendicular to the direction of the wire and the
magnetic field. The symbol $\left\langle ...\right\rangle _{\sec }$ stands
for the averaging across the wire. For the ``flat'' wires made on the basis
of a 2D gas as in Ref.~\cite{Exp}, one has $\left\langle y^{2}\right\rangle
^{1/2}=d/\sqrt{12}$, where $d$ is the width of the wire. For wires with a
circular cross-section $\left\langle y^{2}\right\rangle ^{1/2}=d/4$, where $d
$ is the diameter. Equation (\ref{02}) is written in the limit $X\ll 1$ and $%
r\gg L_{{\rm co}}$. At $X=0,$ Eq.~(\ref{02}) reduces to the well known
result for the wire without a magnetic field \cite{efetov,book}. At small
but finite $X\ll 1 $, the second term in Eq.~(\ref{02}) is small for not
very large $r$ but in the limit $r\rightarrow \infty $ it is always larger
than the first one.

Equation (\ref{02}) shows that a weak magnetic field changes drastically the
tail of wave functions leaving their main body almost unchanged. Even if
there is a correction to the localization length $L_{{\rm co}}$, the second
term in Eq.~(\ref{02}) is more important in the limit of large $r$. In other
words, the characteristic behavior of wave functions at distances $r\ll
r_{X} $ is described by formulae of the orthogonal ensemble, while the tail $%
r\gg r_{X}$ corresponds to the unitary one. Comparing the first and the
second term in Eq. (\ref{02}) one estimates the characteristic distance $%
r_{X}$ 
\begin{equation}
r_{X}\sim -L_{{\rm c}}\ln X\;.  \label{03}
\end{equation}

The fact that any weak magnetic field changes the tail of the wave function
looks quite natural because regions where the amplitude of the wave function
is very small must be very sensitive to external perturbations. The
localization of the wave functions is a result of multiple interference of
waves scattered by impurities. The tails are formed by a coherent
propagation of waves along very long paths, which makes the influence of any
weak magnetic field on the interference extremely important. The separation
into the ``orthogonal'' and ``unitary'' parts loses its sense at $X\sim 1$
corresponding to characteristic magnetic fields determining the crossover
between the orthogonal and unitary ensembles.

Analogous changing of the asymptotic behavior in a magnetic field occurs in
0D (see, e.g.\cite{book}). The level-level correlation function for the
orthogonal ensemble is proportional to the energy $\omega $ in the limit $%
\omega \rightarrow 0$ but any weak magnetic field changes this behavior to
an $\omega ^{2}$ -dependence, which is characteristic for the unitary
ensemble. Of course, the region of the $\omega ^{2}$- dependence is very
narrow at weak magnetic fields.

The factor $X^{2}\ln ^{2}X$ in Eq.~(\ref{02})  is proportional to the square
of the flux through a segment of the wire limited by $r_{X},$ Eq. (\ref{03})
and resembles the Mott law \cite{Mott2,Ber} for the ac conductivity, $\sigma
(\omega )\sim \omega ^{2}\ln ^{2}\omega $. Using the standard perturbation
theory one can understand that $r_{X}$ is the length at which the magnetic
field considerably changes correlations and so, its presence in Eq. (\ref
{2}) is natural.

The final result, Eq.~(\ref{02}), is essentially non-perturbative and its
derivation is not simple. Let us present now the scheme of the computation.

The correlation function $p_{\infty }\left( r\right) $ is extracted from the
low-frequency limit of the density-density correlation function which, in
its turn, is presented as a functional integral over $8\times 8$
supermatrices $Q$ \cite{book} 
\begin{equation}
p_{\infty }\left( r\right) =\lim_{\omega \rightarrow 0}\left( -i\omega
Y_{\omega }\left( r\right) \right) \,  \label{04}
\end{equation}
\[
Y_{\omega }\left( r\right) =-2\pi ^{2}\nu ^{2}\int Q_{24}^{12}\left(
0\right) Q_{42}^{21}\left( r\right) \exp \left( -F\left[ Q\right] \right) DQ 
\]
where $Q_{42}^{12}$ and $Q_{24}^{21}$ are the matrix elements of
supermatrices $Q$. The free energy functional $F\left[ Q\right] $ is written
in 1D in the presence of a magnetic field as 
\begin{equation}
F[Q]=\frac{\pi \nu _{1}}{8}{\rm Str}\int \left[ D_{0}\left( \partial
Q(x)\right) ^{2}+2i\omega \Lambda Q(x)\right] {\rm d}x\,.  \label{1}
\end{equation}
In equation (\ref{1}), we introduced the operator $\partial Q=\nabla
_{x}Q-(ie/\hbar c){\bf A}[Q,\tau _{3}],$ the coordinate $x$ is chosen along
the wire, and the standard notations for the supertrace ${\rm Str}$ and
matrices $\tau _{3}$ and $\Lambda $ are used \cite{book}. Choosing the gauge 
${\bf A}=\left( Hy,0,0\right) $ and using the notations introduced below Eq.
(\ref{02}) one rewrites the functional $F$ as 
\begin{equation}
F\left[ Q\right] =F\left[ Q\right] _{H=0}+X^{2}\left( 32L_{{\rm co}}\right)
^{-1}{\rm Str}\int [Q,\tau _{3}]^{2}\,{\rm d}x\,.  \label{2}
\end{equation}

Calculation of the one-dimensional functional integral in Eq.~(\ref{04}) is
performed using the transfer matrix technique. The Fourier transform of the
correlator $Y_{\omega }$ is written as the integral \cite{book} 
\begin{equation}
Y_{\omega }\left( k\right) =-2\pi ^{2}\nu _{1}\nu \int \Psi\,
Q_{24}^{12}\,(P_{k42}+P_{-k42})\,{\rm d}Q \,  \label{4}
\end{equation}
of the product of the scalar and the matrix functions $\Psi(Q)$ and $P(Q)$,
representing the partition functions of the segments $x<0$ and $x>0$,
respectively. Due to the supersymmetry one has $\int \,\Psi (Q){\rm d}Q=1$.

Discretizing the wire and using Eqs. (\ref{1}, \ref{2}) one derives
recurrence relations for the functions $\Psi $ and $P$ on neighboring sites.
In the continuum limit, these relations yield effective ``Schr\"{o}dinger
equations.'' The corresponding Hamiltonian contains the Laplacian $\triangle
_{Q}$ acting in the space of matrix elements \cite{efetov,book,Zirn}. The
functions $\Psi $ and $P$ have the symmetry of the states with the zero and
the first ``angular momentum,'' respectively. In a schematic form, one has 
\begin{eqnarray}
\left[ -\triangle _{0Q}+X^{2}(\lambda _{1c}^{2}-\lambda _{c}^{2})-\tilde{%
\omega}{\rm Str}(\Lambda Q)\right] \Psi  &=&0\,,  \label{6} \\
\left[ 2ikL_{\rm co}-\triangle _{1Q}+X^{2}(\lambda _{1c}^{2}-\lambda _{c}^{2})-%
\tilde{\omega}{\rm Str}(\Lambda Q)\right] P_{k} &=&Q\,\Psi \,,  \nonumber
\end{eqnarray}
where $\triangle _{0Q}$ and $\triangle _{1Q}$ are the projections of the
Laplacian $\triangle _{Q}$ on the states with the zero and first angular
momentum, and $\tilde{\omega}=i\omega L_{\rm co}^{2}/2D_{0}$.

Equations (\ref{4}, \ref{6}) solving the problem are rather general. A
certain parametrization of $Q$-matrices should be chosen to perform explicit
calculations. As soon as $H\neq 0$, the standard parametrization used in
Refs. \cite{EL,efetov,book} is not convenient, so that we use the ``magnetic
parametrization'' of Refs. \cite{EfAl,book} constructed for studying the
crossover between the orthogonal and unitary ensembles in 0D. However, even
this parametrization, when applied to Eqs. (\ref{6}) leads to extremely
complicated partial differential equations.

Instead of trying to solve these equations exactly, we concentrate on a
study of the limit of weak magnetic fields. The localization length is
determined by the poles of the function $P$ in the $k$-plane \cite{book} in
the region of ``free motion'', faraway from the ``barrier'' given by the
last term in Eq.~(\ref{6}). The position of the poles $k_{n}$ are determined
by the equation ${\cal E}_{n}(k_{n})=0$, where ${\cal E}_{n}(k)$ are
eigenvalues of the operator entering the LHS of Eq.~(\ref{6}) corresponding
to the eigenfunctions $\varphi _{n}(Q)$. According to the general procedure
developed in Ref.~\cite{book}, one has no need to consider solutions at
arbitrary $\omega $. Although the dependence on $\omega $ is necessary for
computing different matrix elements, the poles can be found just by putting $%
\omega =0$. This is natural because the localization properties are
determined for finite samples at $\omega =0$ \cite{Zirn}. Even then, we
cannot find all solutions $P_{k}\left( Q\right) $ and corresponding poles $k$%
. However, in order to determine the function $Y_{\omega }\left( r\right) $
at large distances, we need only one state with the smallest non-zero $%
k_{0}\left( X\right) $. At zero magnetic field, such a state $\varphi
_{0}\left( Q\right) $ is well known \cite{efetov,book} to have a pole at $%
k_{0}\left( 0\right) =\pm i\left( 4L_{{\rm co}}\right) $ $^{-1}$. This
corresponds to the first term in Eq.~(\ref{02}). As soon as one applies a
weak magnetic field, the state is distorted. This effect is not crucial: We
find no corrections in the lowest order either to the localization length or
to the pre-exponential. 

The main result of the present work, namely, the second term in Eq.~(\ref{02}%
), can be obtained without complicated calculations. It turns out that, at
arbitrarily weak magnetic field, an additional state $\varphi_{1}\left(
Q\right) $ with a smaller value $k_{1}=\pm i\left( 8L_{{\rm co}}\right) ^{-1}
$ appears. It is this state that determines the behavior of $p_{\infty
}\left( r\right) $ in the limit $r\rightarrow \infty $. Remarkably, the
value $k_{1}$ does not depend on the magnetic field even in the limit $X
\rightarrow \infty $, when the state $\varphi_{1}$ gives the main
contribution at all distances.

The origin of the states, $\varphi_{0}$ and $\varphi_{1}$, is apparent. An
essential feature of the magnetic parametrization is a finite contribution
arising due to the singularity of the Jacobian at $\lambda_{1c} $, $%
\lambda_{c}\rightarrow 1$, where $\lambda _{1c}$ and $\lambda _{c}$ are
``eigenvalues'' corresponding to the cooperon degrees of freedom. This
singularity is of the type $\left( \lambda _{1c}-\lambda _{c}\right) ^{-2}$
and is usual in the supersymmetry technique. The procedure of regularization
of the singularity has been developed for the 0D cases in Refs.~\cite
{EfAl,book}. In the 1D case, the contribution from the singular term gives
the correlator for the unitary ensemble (in the limit $\lambda_{1c}$, $%
\lambda_{c}\rightarrow 1$ the cooperon degrees of freedom are frozen). In
order to compensate this part at moderate fields, the regular contribution
should contain both the solutions $\varphi_{0}$ and $\varphi_{1}$. At $%
X\rightarrow 0$, the solution $\varphi_{1}$ exactly compensates the singular
term, so that the orthogonal part is given by $\varphi_{0}$ only.

Having disregarded the term with the external frequency $\omega $, we arrive
at a Hamiltonian with separated cooperon and diffuson variables. However, it
is still too complicated since essential values of $\lambda _{c}$ and $%
\lambda _{1c}$ are such that $\lambda _{1c}$, $\lambda _{c}\simeq 1$. This
difficulty is avoided by considering the derivative over the magnetic field, 
$Y_{X}^{\prime }$. For the quantity $Y_{X}^{\prime }$, in contrast, large $%
\lambda _{1c}$ and $\lambda _{1d}$ 
become important, which substantially simplifies the ``Hamiltonian'' 
\[
{\cal H}=-\lambda _{1d}^{2}\partial _{\lambda _{1d}}^{2}-\lambda
_{1c}^{2}\partial _{\lambda _{1c}}^{2}+2\lambda _{1c}\partial _{\lambda
_{1c}}+X^{2}\lambda _{1c}^{2}\,.
\]

The solution $\Psi $ decaying at $\lambda _{1c}\rightarrow \infty $ equals  
\begin{equation}
\Psi =(1+X\lambda _{1c})\exp (-X\lambda _{1c})\,,  \label{psi}
\end{equation}
and satisfies the boundary condition $\Psi (0)=1$. The main contribution to
the correlator $Y_{X}^{\prime }$ comes from the region $1\ll \lambda
_{1c}\ll 1/X$ within logarithmic accuracy and the function $\Psi $, Eq.~(%
\ref{psi}), takes the form $\Psi =1-X^{2}\lambda _{1c}^{2}/2$. 

There are two terms in the expression for $Y^{\prime }\sim \int {\rm d}%
Q\,Q_{24}^{12}\,(\Psi _{X}^{\prime }P+\Psi P_{X}^{\prime })$, each of them
having a different localization length. The function $P$ determined from 
\begin{equation}
(-\lambda _{1d}^{2}\partial _{\lambda _{1d}}^{2}-\lambda _{1c}^{2}\partial
_{\lambda _{1c}}^{2}+2\lambda _{1c}\partial _{\lambda _{1c}}+2ikL_{\rm co})P=Q
\label{p0}
\end{equation}
is given by $P=\frac{1}{2}Q/(1+ikL_{\rm co})$. This leads to a fast decaying
contribution to the function $Y$. The main contribution at large distances
comes from the function $P_{X}^{\prime }$ satisfying the equation 
\begin{equation}
({\cal H}+2ikL_{\rm co})P_{X}^{\prime }\simeq -X\lambda _{1c}^{2}Q\,.  \label{p1}
\end{equation}
Recalling that $Q\sim \lambda _{1c}\lambda _{1d}$, we represent $%
P_{X}^{\prime }$ in the form $P_{X}^{\prime }=XQ\lambda _{1c}^{2}f(\lambda
_{1d})$ and obtain 
\begin{equation}
(-\lambda _{1d}^{2}\partial _{\lambda _{1d}}^{2}-2\lambda _{1d}\partial
_{\lambda _{1d}}+2ikL_{\rm co})f\simeq 1\,.  \label{f}
\end{equation}
This expression is essentially the same as the one for the unitary ensemble,
yielding the localization length $L_{cu}$. The pre-exponential of the
correlator $Y_{X}^{\prime }$ is most easily estimated for $k=0$ when $%
f=-\ln \lambda _{1d}$. Estimating the derivative of Eq.~(\ref{4}%
) 
\begin{equation}
Y_{X}^{\prime }\sim X\int^{1/X}{\rm d}\lambda _{1c}\int^{1/\lambda _{1c}}%
{\rm d}\lambda _{1d}\ln \lambda _{1d}\sim X\ln ^{2}X\,,  \label{corr}
\end{equation}
we integrate the result over $X$ and use the fact that the correlator $%
p_{\infty }(r)$ at vanishing magnetic field must coincide with the one of
the orthogonal ensemble. This leads us finally to Eq.~(\ref{02}).

It is relevant to note that in our analysis we considered the limit when $%
\omega \rightarrow 0$ but the magnetic field $H$ remains finite. In
principle, this limit differs from the limit $H=0$, $\omega \rightarrow 0$
considered in Refs. \cite{EL,efetov}. However, for the correlation function $%
p_{\infty }\left( r\right) $, Eq. (\ref{0}), the limit $H\rightarrow 0$ must
give the same results as those obtained for $H=0$.

Although our analysis was performed for very large distances $r$, it allows
us to believe that the entire range of fields between the orthogonal and
unitary ensembles is governed by an interplay between the two localization
lengths. On the other hand, one should keep in mind that Eq.~(\ref{02}) is
not exact in the sense that it remains to be seen how the localization
lengths and the pre-exponentials change at arbitrary $X$.

The localization length $L_{c}$ enters directly such a physical observable
as the hopping conductivity \cite{kurk,larkin}. However, previous theories
as well as the interpretation of the experiment of Ref. \cite{Exp} assumed
the presence of only one characteristic length. For the hopping conductivity
in quasi-1D samples, the accounting for the two localization lengths can be
performed using Eqs.~(\ref{01}, \ref{03}). It is not difficult to understand
that the activation law, Eq.~(\ref{01}), with $L_{c}=L_{{\rm co}}$, holds
for temperatures $\left( \nu _{1}r_{X}\right) ^{-1}<T<\left( \nu
_{1}L_{c}\right) ^{-1}$. In this regime, the change of the asymptotic
behavior of wave functions at very large distances is not essential.
Nevertheless, the far asymptotics is extremely important for calculation of
the conductivity at lower temperatures $T<\left( \nu _{1}r_{X}\right) ^{-1}$%
. Repeating arguments of Refs. \cite{kurk,larkin} one comes to the
conclusion that Eqs. (\ref{01}) can be used also for the description of the
latter regime provided the temperature $T_{0}$ is replaced by $T_{0}/2$.
This means that, experimentally, decreasing temperature should lead to a
crossover from the activation behavior, Eq. (\ref{01}), to another
activation behavior with the characteristic temperature $T_{0}/2$.
Measurements at lower temperatures than those used in Ref. \cite{Exp} might
help to check our predictions.

In conclusion, we have demonstrated that the far tail of the density-density
correlator in disordered wires at arbitrary weak magnetic field decays with
the localization length of the unitary ensemble $L_{{\rm cu}}$, which is
double as large as the length $L_{{\rm co}}$ of the main part. This means
that there is no one-scale crossover between the limits $H=0$ and $H=\infty $%
. An arbitrary weak magnetic field drastically changes the decay of wave
functions: the initial decrease with the localization length $L_{{\rm co}}$
persists only up to the distances $r_{X}\sim -L_{c}\ln X$ but then it is
followed by the decay with the length $L_{{\rm cu}}$. An increase of the
magnetic field leads to a respective decrease of $r_{X}$ so that finally, at 
$X\sim 1$, wave functions decay everywhere with the length $L_{{\rm cu}}$.
The behavior found can manifest itself in the hopping conductivity, which
might be the simplest way to test our predictions experimentally.

\end{multicols} 

\end{document}